# *In-chip* microstructures and photonic devices fabricated by nonlinear laser lithography deep inside silicon

Onur Tokel[1], Ahmet Turnalı[2], Ghaith Makey[1], Parviz Elahi[1], Serim Ilday[1], Tahir Çolakoğlu[3], Emre Ergeçen[4], Özgün Yavuz[2], René Hübner[5], Mona Zolfaghari Borra[3,6], Ihor Pavlov[1], Alpan Bek[3,6,7], Raşit Turan[3,6,7], Denizhan Koray Kesim[2], Serhat Tozburun[8,†], F. Ömer Ilday[1,2,‡]

[1] Department of Physics, Bilkent University, Ankara, 06800, Turkey

[2] Department of Electrical and Electronics Engineering, Bilkent University, Ankara, 06800, Turkey

[3] The Center for Solar Energy Research and Applications, Middle East Technical University, Ankara, 06800, Turkey

[4] Department of Electrical Engineering and Computer Science, Massachusetts Institute of Technology, Cambridge, Massachusetts, 02139, USA

[5] Helmholtz-Zentrum Dresden - Rossendorf, Institute of Ion Beam Physics and Materials Research, Bautzner Landstraße 400, 01328 Dresden, Germany

[6] Micro and Nanotechnology Graduate Program, Middle East Technical University, Ankara, 06800, Turkey

[7] Department of Physics, Middle East Technical University, Ankara, 06800, Turkey

[8] Harvard Medical School, Boston, MA, 02115, USA

[†] Present address: Izmir International Biomedicine and Genome Institute, Izmir, 35340, Turkey

[‡] To whom correspondence should be addressed: ilday@bilkent.edu.tr

**Silicon is an excellent material for microelectronics and integrated photonics[1-3] with untapped potential for mid-IR optics[4]. Despite broad recognition of the importance of the third dimension[5,6], current lithography methods do not allow fabrication of photonic devices and functional microelements directly inside silicon chips. Even relatively simple curved geometries cannot be realized with techniques like reactive ion etching. Embedded optical elements, like in glass, electronic devices and better electronic-photonic integration are lacking[7,8]. Here, we demonstrate laser-based fabrication of complex 3D structures deep inside silicon using 1 µm-sized dots and rod-like structures of adjustable length as basic building blocks. The laser-modified Si has a different optical index than unmodified parts, which enables numerous photonic devices. Optionally, these parts are chemically etched to produce desired 3D shapes. We exemplify a plethora of subsurface, *i.e.*, "*in-chip*" microstructures for microfluidic cooling of chips, vias, MEMS, photovoltaic applications and photonic devices that match or surpass the corresponding state-of-the-art device performances.**

We begin by discussing point-like modification of Si at laser focus, based on an approach we first demonstrated[9]. We use a custom-developed nanosecond-pulsed fibre laser[10] operating at a central wavelength of 1.55 μm, where Si is transparent. Using a 40× objective to tightly focus the beam, we trigger a set of light-matter interaction mechanisms simultaneously (Supplementary Information section 3), which leads to well-controlled and repeatable beam collapse (Fig. 1a). When only a single pulse is applied, this results in permanent modification of the crystal structure (Supplementary Information section 1) within a roughly 1 μm-wide spherical region (Fig. 1b). By repositioning the beam and sending a single pulse to each new position, any desired 3D structure, like the 1 μm-thick, 1 mm-long helix in Fig. 1c, can be created, point by point. However, like all 3D fabrication methods, there is an inherent trade-off between resolution, speed and processing volume[11]. We provide numerous examples of structures and devices in Figures 2 and 3, which occupy volumes ranging between 0.1-10 $mm^3$. Point-by-point processing such macroscopic volumes with 1-μm resolution requires order of $10^{8-10}$ beam positioning steps, which corresponds to hundreds to thousands of hours of processing, limited by the speed of modern beam positioning equipment. We circumvent this problem by augmenting point-by-point processing with creation of self-organized rod-like structures forming along the beam propagation direction. This works well in practice since most 3D structures of interest can be broken down to rods of various lengths (rods grow in length with number of incident pulses). If the structure indeed requires finer features, these are created by additional point-by-point processing. This way, the fabrication time is reduced from thousands of hours to mere minutes.

In order to trigger the self-organized pulse-to-pulse growth of the rod-like structures, we invoke nonlinear feedback mechanisms[12-15] arising from interaction of infrared laser pulses inside Si: (i) Competing thermal nonlinearity- and free carrier-induced (FCI) refractive index changes due to counter-propagating laser beams, (ii) a self-limiting, non-local feedback loop arising from permanent local modification of Si at beam focus, which, in turn, iteratively shifts the focal position of the beam, resulting in rod-like structures, the lengths of which are controlled by the number of incident pulses. The beam, focussed by a lens, passes through the Si chip; about a third of it returns from the back surface of the chip due to Fresnel reflection and counter-propagates with respect to the incident beam (Fig. 1a). Near its focal point, positioned at a desired point inside the chip, the beam collapses due to nonlinear effects and modifies Si. When additional pulses are sent, their focal points iteratively shift along the optical ($z$) axis as each pulse further modifies the material, changing the optical path of the next pulse (Fig. 1a). The beam width is unaltered, so the structured region progressively gets longer without increasing in width and its aspect ratio can reach far beyond the diffraction limit. By repositioning the beam and sending a controlled number of pulses, we create elongated structures of controllable length (Fig. 1g) at desired locations within the chip, which are combined to construct complex structures (Fig. 1c, 1d). After this single-step procedure, the Si chip has only internal refractive index modification, which is used to demonstrate a variety of optical elements, *e.g.*, lenses, holograms, waveguides. As an optional second step, we can selectively remove the modified regions by chemical etching to enable full 3D sculpting of the entire chip. For instance, the fabrication of Penrose Stairs[16] is illustrated in Fig. 1e.

We will explain the self-organisation physics governing the creation of the structures through a toy model (Supplementary Information section 2). Wave propagation is governed by the nonlinear paraxial equation and the heat equation. This equation system is mathematically equivalent to the Newton-Schrödinger system[17]. Two competing diffractive effects dominate beam propagation: Thermally induced change of index of refraction ($\Delta n_{\text{therm}}$) and free carrier-induced (FCI) refractive index change ($\Delta n_{\text{FCI}}$), which are similar to concatenated converging and diverging lenses with focal lengths, $f_{\text{therm}}$ and $f_{\text{FCI}}$, respectively. Modification of Si occurs if the beam self-focuses and collapses (the first feedback mechanism), which requires thermal lensing to be stronger than diffraction due to FCI effect ($f_{\text{therm}} < |f_{\text{FCI}}|$). The incident and reflected beams are individually too weak to modify the material ($f_{\text{therm}} > |f_{\text{FCI}}|$), but when overlapped, they cooperate to permanently modify the crystal structure (Fig. 1f). This scenario is verified by numerical solutions of beam propagation based on the split-step Fourier method (Supplementary Information section 3), which show that shielding due to negative $\Delta n_{\text{FCI}}$ precludes subsurface modification for a single beam, consistent with experiments. In contrast, thermal lensing is enhanced when the beams are coupled (Fig. 1a, 1f) and the beam collapses to induce permanent refractive index changes in Si. When more than one pulse is incident, the second feedback mechanism is activated: Each pulse locally modifies silicon; the resulting change in refraction index shifts the focal point for the next pulse similar to the moving focus model of self-focusing[18]. Focal position of the second pulse, $l_2$, is given by $\frac{1}{l_2} = \frac{1}{f_{\text{FCI}}} + \frac{1}{f_{\text{therm}}} + \frac{1}{l_1}$, where $l_1$ is the focal position of the first pulse. The location of the modification caused by the $n^{\text{th}}$ pulse (n>1), $l_{\text{n}}$, is found from $\frac{1}{l_n} = (n-1)\left(\frac{1}{f_{\text{FCI}}} + \frac{1}{f_{\text{therm}}}\right) + \frac{1}{l_1}$. The total length is then $l_{\text{total}} = l_1 - l_n = \frac{l_1}{1+\xi/(n-1)}$, where $\xi = \frac{f_{\text{FCI}} f_{\text{therm}}}{l_1(f_{\text{FCI}} + f_{\text{therm}})}$ is a constant. This result reveals that elongation of the structures is self-limiting, stopping before reaching the chip surface (Fig. 1g).

Now, we discuss the features and limitations of creating general 3D structures out of the rod-like building blocks. First, we verify that continuous, wall-like structures can be created by translating the beam in a plane perpendicular to the beam propagation direction (*z*-axis) (Fig. 1d and Movie S1). Cross-sectional microscope images show that the length of the structures range from the Rayleigh length (~20 µm, Fig. 1g) to ~1 mm (limited by the thickness of the chip), while always retaining a width of ~1 µm (Fig. 1d). The corresponding aspect ratio approaches 1000, which is only possible due to the iterative formation mechanism. Next, we verify that the structures can cross over each other without distortion, as evidenced by the mesh structure shown in Fig. 1d. We also confirm that curvilinear structures can be formed (see the spiral in Figure S14). None of these features were obvious, *a priori*, due to the self-organised nature of the formation mechanism. There is no fundamental limit to the area over which the structures can be created. Similarly, doping of silicon is not a limitation; we can create them in both intrinsic and doped Si. A complete set of experiments is provided for n- and p-doped Si chips, where the effects of polarisation, scanning angle and direction are analysed (Supplementary Information section 4). Furthermore, we do not observe any asymmetry resulting from the scanning-direction or polarisation of the beam, which is not surprising since Si is a centrosymmetric material[19].

We first discuss demonstrations based on purely internal index modification of the Si chip, followed by those where the chip is selectively etched to reveal a desired 3D architecture. There is much interest in using silicon as a material for mid-IR applications[4]. The capability to locally modify the index of refraction in a well-controlled manner affords unprecedented spatial phase control over optical beams. As an example, we demonstrate the first in-chip Fresnel zone plate (FZP) lens (Fig. 2a). The focal length, $f$, is determined by $R_k = \sqrt{kf\lambda + (k\lambda/2)^2}$, where $R_k$ is the radius of $k^{\text{th}}$ zone and $\lambda$ is the wavelength. A FZP is designed with $R_{20} = 1.5$ mm, $f = 7.2$ cm at 1.55 µm. The performance is evaluated to give $f = 7.3$ cm, $M^2 = 1.46$ (Fig. 2a) with 15% efficiency, which exceeds that of single-layer FZP lenses created in glass.

Additional applications of creating a refractive index distribution inside silicon include wavefront structuring (*20-22*), flat optics (*23*), adaptive optics (*24*), Si photonics (*1*) and better electronic-photonic integration (*25*). We showcase this capability through creation of 2D Fourier computer generated holograms (CGHs) for binary (Fig. 2b) and greyscale (Fig. 2c) images, and Fresnel (Fig. 2d) CGHs for 3D images, all designed using a customised adaptive-additive iterative Fourier transform algorithm (Supplementary Information section 5). The experimentally reconstructed binary hologram, composed of 360 × 360, 10-µm pixels, is shown in Fig. 2b. Reproduction of high-frequency spatial components in holography is notoriously difficult (*25*). The fact that they are reproduced in silicon with high quality is a confirmation of the good control we have over index modifications (Fig. 2b). Greyscale images require even more advanced wavefront structuring, which we demonstrate through a 600 × 600, 10-µm pixel hologram (Fig. 2c). The pixel count and image reconstruction quality of our in-chip holograms surpass those of state-of-the-art nanophotonic phase arrays (64 × 64, 9-µm pixels)[21] and metamaterial phase holograms (150 × 150, 5-µm pixels)[25]. We also note that Si is CMOS compatible and lossless over a broad spectral range, including telecommunication wavelengths and much of the mid-IR, unlike metamaterials[26]. Finally, we demonstrate holography of a 3D image using a Fresnel hologram (800 × 600, 10-μm pixels), which projects rotating rectangles at consecutive planes (Fig. 2d and Movie S2). This is achieved by first combining a stack of Fourier holograms, each generating a 2D slice of an intended 3D object, and then superimposing them with FZPs, which shift each image to its corresponding focal plane (Supplementary Information section 5). This construction itself is novel, whereby the phase-type Fresnel hologram can be generated without directly using the Fresnel equation. To assess hologram efficiency[27], $R$, gratings were written in Si, with $R = \frac{4F\sin^2(\Delta\phi)}{\pi^2(1-F+F\cos^2(\Delta\phi))}$, where $\Delta\phi$ is the phase modulation and $F$ is the filling factor. The measured $R$ of ~150% corresponds to $\Delta\phi = 0.69\pi \pm 0.04\pi$, which compares well with the directly measured $\Delta\phi = 0.69\pi \pm 0.01\pi$.

As final applications of purely optical modification of silicon, we demonstrate formation of an optical waveguide (Fig. 2e) and in-chip information storage (Fig. 2f). For the former, we created a 2.5 mm-long tubular waveguide structure since the laser-processed regions have a depressed index of refraction. The radius of the core region is 10 μm. Far-field intensity profile of the beam obtained at the exit of the Si chip is shown in Fig. 2e, where an identical measurement of the beam traversing a region without a waveguide is also provided as control. For the latter, we create multi-level (more than 500 levels are theoretically possible assuming a 1 mm-thick chip; as proof of principle, 25 were written, 7 of which can be directly seen in Movie S3) buried structures arranged in dot or barcode format. Next, we created a two-level barcode, where the levels spell "UFO" and "LAB" (Fig. 2f). We limited ourselves to two levels for ease of readout, which was achieved by optical coherence tomography. Furthermore, laser-written structures could be erased, following exposure to high temperatures (1100 °C) in an oven for 2 hours. Afterwards, the same digital readout protocol that correctly detected 96% of the written dots false-detected only 3% of erased dots (Supplementary Information section 4), highlighting the prospects for erasable and rewritable information storage and holography. This process may likely be simplified and rendered spatially selective by using continuous-wave laser heating instead of an oven.

The second class of applications is based on selective chemical etching of the optically processed chip for truly 3D sculpting (Fig. 3). Thanks to the vastly higher etching rates for modified parts of silicon compared to unmodified parts, we are able to create extremely regular arrays of sharp-edged and high-aspect-ratio (20 μm × 30 μm × 500 μm) pillars within minutes (Fig. 3a and Supplementary Information section 6). The average surface roughness of the etched structures has been measured to be ~20 nm using atomic force microscopy. This is much smaller than our smallest feature size of 1 μm and the IR wavelengths for which Si is transparent (>1 μm). Among the numerous possible applications, there is the possibility to create buried microfluidic channels (Fig. 3b). Pumping coolants directly through microfluidic channels written inside chips is being considered as a possible remedy to overheating of microprocessor chips[28], which is a main limitation to increasing clock rates[29]. As a proof-of-principle experiment, we passed cold water through a silicon chip, which decreased the surface temperature by 4 °C within a few seconds (Fig. 3c and Movie S4). As another potential application of 3D sculpting to electronics, we demonstrate through-Si vias[30] of controllable depth, including those that cut across the entire chip (Fig. 3d). Further, we show 3D cantilever-like structures that could find use in MEMS applications (Fig. 3e). We also show controlled slicing of a chip into ~30 µm-thick plates (Fig. 3f). To this end, planar sections inside the chip are raster-scanned with laser pulses, followed by chemical etching, resulting in mechanical separation of the chip along the processed planes. The sliced plates can be used for, *e.g.*, photovoltaic cell fabrication, reducing the required amount of Si material by an order of magnitude with potential for major cost reduction. As a final illustration of the power of 3D sculpturing, we created a microscale silicon city, *Siliconopolis*, with representations of buildings, trees, a beach and a sea in Fig. 3g.

We have outlined a general metholodogy and the various examples demonstrated here are far from constituting an exhaustive list. Additional possibilities include creation of in-chip microlens arrays, microresonators, modulators for multi-level electronic-photonic integration and rewritable holograms or information storage.

# Methods

## Experimental setup

We used optical pulses with a central wavelength of 1.55 μm, duration of 5 ns and repetition rate of 150 kHz (Supplementary Information section 7). The laser system was coupled to a processing station with precise alignment, power and polarisation control. For single-point modification, a 40× objective with NA of 0.75 was used. For rod-like structures, the laser was focused to ~3 μm ($f$ = 4.5 mm, NA = 0.55) after the back surface of a Si sample, which produced the counter-propagating beam. The samples were translated with a 3-axis high-resolution computer-controlled stage (Aerotech, ANT130-XY, ANT95-L-Z). The pulse energy was controlled with a half-wave plate between two polarising beam splitters (PBS) operating as an attenuator, while the polarisation after the latter PBS was linear and in the plane of the sample surface. A second half-wave plate was used to control the polarisation angle. Pulse energies were in the range of 8 μJ and 15-20 μJ for single-point and rod-like structures, respectively. Experiments were performed at room temperature and in ambient atmosphere.

## Optical coherence tomography and IR microscopy

In order validate the 3D information encoding/decoding capabilities in Si, we used a second-generation optical coherence tomography system called optical frequency-domain imaging (OFDI). The OFDI system uses a polygon filter-based wavelength-swept laser source operating at an axial scan rate of 50 kHz. The sweeping range is over 120 nm at a centre wavelength of 1.3 μm, providing 6-μm axial resolution in air (~1.8 μm in Si). The optical interference signals were digitalised and directly transferred to a hard-drive storage array by a two-channel 100 MS/s data acquisition board (Signatec, Inc). Approximately 10% of this acquired data was employed and processed on the CPU for real-time structural image feedback. We used a homemade scanner microscope comprising a two-axis galvanometre mirror system (GVS112, Thorlabs, Inc.) and a 10× scan lens (LSM02, Thorlabs, Inc.) for barcode decoding. The barcodes use international symbology UPC, code 128.

A homemade IR microscope was used to evaluate the subsurface structures, which incorporates a broad-spectrum halogen lamp as light source, and an EMCCD camera (Andor, Luca S) as detector. The IR microscope was operated in the transmission mode with a 20× objective (Nikon, 0.45 NA 20×). Laser-written samples were studied with a reflection mode optical microscope (Nikon, 0.6 NA, 40×) and also with SEM imaging, confirming that no photo-induced modifications on either side of the wafer. In addition, etching results provided additional confirmation of undamaged surfaces after laser writing.

**Sample preparation, positioning and scanning**

Double-side-polished, <100>-cut, p-type Si samples (boron doped, 1 Ω.cm and 15 kΩ.cm) and n-type (phosphorus doped, 1 Ω.cm) were used (Siegert Wafer, GmbH). 500 μm - 1.3 mm thick wafers were diced in 15 mm by 30 mm dimensions. First, a piranha solution was applied to samples (2 minutes). Then, they were cleaned in acetone (5 minutes), ethanol (5 minutes) and deionised water (5 minutes) and finally dried with nitrogen. Si samples were then held in position with neodymium magnets. Precise sample alignment with respect to the laser was found to be crucial. This was accomplished by using interference of multiple reflections from the sample's polished surfaces, which was monitored with an IR sensitive camera. The scheme allowed alignment within a few micrometres throughout the scanning range, and also operated as a proxy for *in-situ* imaging. The scans were performed in a plane perpendicular to the laser propagation direction at speeds in the 0.2-7 mm/s range.

**Chemical etching procedure**

Chemical etching is applied to optically processed Si samples after the following preparation steps. First, the samples were lapped down 150-250 μm from the surface (Allied High MultiPrep). Consecutive lapping steps were applied with progressively smaller diamond-particle decorated films (with sizes of 35 μm, 9 μm, 6 μm and 0.5 μm). Then, further polishing is performed using an alumina suspension (particle size of 50 nm), until a mirror-like surface is obtained. In order to prevent contamination, the polished surface is cleaned with the Radio Corporation of America (RCA) clean procedure (RCA-1 and RCA-2 steps). The samples were then exposed to the etching solution for a few minutes, in which the molar concentrations of the components, $Cu(NO_3)_2$, HF, $HNO_3$ and $CH_3COOH$ were 0.05 M, 10 M, 4 M and 3.5 M, respectively (see Supplementary Information section 6). Finally, the samples were rinsed with deionized water and dried with nitrogen flow. All chemicals were supplied from Merck Chemical Company.

# Bibliography

1. J. Leuthold, C. Koos, W. Freude, Nonlinear silicon photonics. *Nature Photon.* **4**, 535–544 (2010).
2. F. Priolo, T. Gregorkiewicz, M. Galli, T. F. Krauss, Silicon nanostructures for photonics and photovoltaics. *Nature Nanotech.* **9**, 19–32 (2014).
3. A. E.-J. Lim *et al.*, Review of silicon photonics foundry efforts. *IEEE J. Select. Quant. Electron.* **20**, 405–416 (2014).
4. R. Soref, Mid-infrared photonics in silicon and germanium. *Nature Photon.* **4**, 495–497 (2010).
5. P. G. Emma, E. Kursun, Is 3D chip technology the next growth engine for performance improvement? *IBM J. Res. Dev.* **52**, 541–552 (2008).
6. H. Wong, S. Salahuddin, Memory leads the way to better computing. *Nature Nanotech.* **10**, 191 (2015).
7. N. Sherwood-Droz, M. Lipson, Scalable 3D dense integration of photonics on bulk silicon. *Opt. Express*. **19**, 17758–17765 (2011).
8. Z. Wang, 3-D integration and through-silicon vias in MEMS and microsensors. *Journal of Microelectromechanical Systems*. **24**, 1211–1244 (2015).
9. I. A. Pavlov, E. Dulgergil, E. Ilbey, F. Ö. Ilday, (Optical Society of America, 2012), p. CTu2M.5.
10. I. Pavlov, E. Dülgergil, E. Ilbey, F. O. Ilday, Diffraction-limited, 10-W, 5-ns, 100-kHz, all-fiber laser at 1.55 μm. *Opt. Lett.* **39**, 2695–2698 (2014).
11. R. L. Truby, J. A. Lewis, Printing soft matter in three dimensions. *Nature*. **540**, 371–378, 2016.
12. B. Oktem *et al.*, Nonlinear laser lithography for indefinitely large-area nanostructuring with femtosecond pulses. *Nature Photon.* **7**, 897–901 (2013).
13. F. T. Arecchi, S. Boccaletti, P. Ramazza, Pattern formation and competition in nonlinear optics. *Phys. Rep.* **318**, 1–83 (1999).
14. C. Kerse *et al.*, Ablation-cooled material removal with ultrafast bursts of pulses. *Nature*. **537**, 84–88 (2016).
15. S. Ilday *et al.*, Multiscale self-assembly of silicon quantum dots into an anisotropic three-dimensional random network. *Nano Lett.* **16**, 1942–1948 (2016).
16. L. S. Penrose, R. Penrose, Impossible objects: a special type of visual illusion. *British Journal of Psychology*. **49**, 31–33 (1958).
17. R. Bekenstein, R. Schley, M. Mutzafi, C. Rotschild, M. Segev, Optical simulations of gravitational effects in the Newton-Schrodinger system. *Nature Phys.* **11**, 872–878 (2015).
18. A. Brodeur *et al.*, Moving focus in the propagation of ultrashort laser pulses in air. *Opt. Lett.* **22**, 304–306 (1997).
19. W. Yang, P. G. Kazansky, Y. P. Svirko, Non-reciprocal ultrafast laser writing. *Nature Photon.* **2**, 99–104 (2008).
20. A. V. Kildishev, A. Boltasseva, V. M. Shalaev, Planar photonics with metasurfaces. *Science*. **339**, 1232009 (2013).
21. J. Sun, E. Timurdogan, A. Yaacobi, E. S. Hosseini, M. R. Watts, Large-scale nanophotonic phased array. *Nature*. **493**, 195–199 (2013).
22. G. Zheng *et al.*, Metasurface holograms reaching 80% efficiency. *Nature Nanotech.* **10**, 308–312 (2015).
23. N. Yu, F. Capasso, Flat optics with designer metasurfaces. *Nature Mater.* **13**, 139–150 (2014).
24. Q. Wang *et al.*, Optically reconfigurable metasurfaces and photonic devices based on phase change materials. *Nature Photon.* **10**, 60–65 (2016).
25. S. Larouche, Y.-J. Tsai, T. Tyler, N. M. Jokerst, D. R. Smith, Infrared metamaterial phase holograms. *Nature Mater.* **11**, 450–454 (2012).
26. C. Wu *et al.*, Spectrally selective chiral silicon metasurfaces based on infrared Fano resonances. *Nature Comm.* **5**, 3892 (2014).
27. X. Li *et al.*, Athermally photoreduced graphene oxides for three-dimensional holographic images. *Nature Comm.* **6**, 6984 (2015).

28. B. Dang, M. S. Bakir, D. C. Sekar, C. R. King Jr, J. D. Meindl, Integrated microfluidic cooling and interconnects for 2D and 3D chips. *IEEE Trans. Adv. Pack.* **33**, 79–87 (2010).
29. P. Ball, Feeling the heat. *Nature*. **492**, 174–176 (2012).
30. M. Motoyoshi, Through-silicon via (TSV). *Proc. IEEE*. **97**, 43–48 (2009).

# Figure Captions

**Figure 1 Creation of 3D structures inside silicon. (a)** The schematic shows laser pulses incident on a Si chip. The pulse collapses and modifies local Si crystal structure, which constitutes a building block for more complex structures. Consecutive laser pulses focus to shifted positions, axially elongating the structured region. The inset shows a scanning electron microscope (SEM) image of a rod-like structure in Si. **(b)** IR image of an array of 1 μm-sized voxels each created by a single laser pulse. **(c)** IR image of a 1 μm-thick, 1 mm-long helix exemplifies point-by-point fabrication of an in-chip 3D structure. **(d)** Various 3D structures can alternatively be formed using rod-like structures as building blocks. These 1-μm wide structures can range from 20 μm to hundreds of micrometres along the laser propagation direction (*z*-axis). SEM (for top surface and cross-section) and infrared (IR) transmission microscope (for subsurface) images correspond to colour-labelled planar sections of the 3D schematic. The subsurface mesh is directly revealed in the IR image, whereas the top (shown) and bottom (not shown) surfaces remain unscathed. **(e)** Illustration summarises single-step selective chemical etching for removal of the laser-processed volumes. The SEM image shows experimental realisation of Penrose Stairs (*16*). **(f)** Simulation results show self-induced refractive index change, comprised of thermal nonlinearity (positive $\Delta n_{\mathrm{thermal}}$) and free carrier induced refractive index change (negative $\Delta n_{\mathrm{FCI}}$) for a single pulse (black curve). The red curve shows the self-induced refractive index change for a single beam focused in Si. The blue curve shows the index change for the two counter-propagating beams. The thermal nonlinearity is dominant in the latter, enabling beam collapse and modification in Si. **(g)** Prediction of our analytic model for structure elongation with each pulse is compared with experiments. Data points represented with red circles were created using a lens with NA of 0.55 and the isolated data point represented by a blue diamond was created using an objective with NA of 0.75. In all figures, *z*-axis is the optical axis.

**Figure 2 Functional in-chip optical elements and information storage. (a)** Composite IR transmission microscope image of a subsurface Fresnel-zone-plate lens (top). The scale bar is 400 μm. Inset shows a close-up view. Theoretical and experimental beam profiles during focusing are shown (bottom). **(b)** Simulation of a binary image with high spatial components, from a hologram of 356 × 356 pixels **(i)**. Experimental reconstruction from the hologram written in Si **(ii)**. **(c)** Simulation of a grey-scale image of Mona Lisa, from a hologram of 600 × 600 pixels **(i)**. Experimental reconstruction from the hologram written in Si **(ii)**. **(d)** Illustration shows 3D holography of a rotating rectangle at four consecutive planes (top). The Fresnel-type hologram written in Si is comprised of 800 × 600 pixels. Experimentally reconstructed images of the projected rectangles are shown below. All holograms have 10-μm sized square pixels. **(e)** Measured

far-field intensity profiles corresponding to the beam traversing a region without a waveguide (control) and a laser-written waveguide. **(f)** Illustration of multi-level information storage and their experimentally measured images in Si. Level 1 barcode spells UFO, and Level 2 barcode spells LAB.

**Figure 3 Sculpting of 3D arbitrary micro-architectures. (a)** SEM image of large-area covering, high-aspect-ratio micropillars revealed after chemical etching. The inset shows a close-up view of the etched micropillar array. The pillars have 20 × 30 μm top surfaces, and their heights extend ~500 μm along the laser propagation direction. **(b)** SEM image of a sidewall of a Si chip, showing embedded microchannels. Inset shows laser scanning microscopy of the channels, which penetrate hundreds of micrometres into the chip. **(c)** The illustration shows an 8 mm-long microfluidic channel carrying cooling water into a chip. Thermal camera images **(i)** before and **(ii)** after passing the water shows that the surface temperature of the chip decreases by 4°C within a few seconds. **(d)** SEM image of entrance of through-Si-vias that cut across the entire chip. **(e)** SEM image of 3D cantilever-like structures. **(f)** SEM image of controlled slicing of a chip into ~30 μm-thick plates. **(g)** An artificially coloured view of a micro-cityscape created out of silicon.

## Movie Captions

**Movie S1 Infrared transmission microscopy of subsurface structures in Si.** IR microscopy of parallel subsurface structures and their projections are shown in the *x*-*y* plane at different depths. The structures elongate hundreds of micrometres along the optical (*z*) axis. The unaltered surface plane is also shown, with shadows of subsurface structures visible through transmission.

**Movie S2 3D projection of a subsurface Fresnel hologram.** The projected images of a Fresnel hologram (800 × 600, 10-µm pixels) create rotating rectangles at different planes, realising 3D projection. Optically reconstructed images form on a screen, which translates at constant speed relative to the camera.

**Movie S3 Multi-level structures in Si.** An array of single pulse modifications was used to create 25-levels along the optical (*z*) axis. IR microscopy of 7 levels are shown. The focal plane of the imaging objective is translated along *z* axis, such that a few levels are visible at each depth.

**Movie S4 Silicon in-chip cooling with embedded microfluidic channels.** Microfluidic cooling of Si is realised by passing cold water through an embedded microchannel. The thermal camera recording is shown on the left, and simultaneously recorded optical camera video is shown on the right.

## Acknowledgments

This work was supported partially by the European Research Council (ERC) Consolidator Grant ERC-617521 NLL, EU Marie Curie Fellowship 660769 SMILE, and TÜBITAK under project 113M930. We gratefully acknowledge support by the Structural Characterization Facilities at IBC of the HZDR. We thank H. Volkan Hünerli for discussions on the chemical procedure.

## Author contributions

A.T. and O.T. contributed equally. O.T. and F.Ö.I designed the research and interpreted the results with help from S.I. Experiments were performed by A.T., O.T., G.M., Ö.Y. The customised laser was built by I.P. Analytical model was developed by P.E., O.T. and F.Ö.I. Numerical simulations were performed by A.T., O.T., E.E. Chemical etching developed by T.Ç, M.Z.B., A.B., R.T. Material analyses were performed by R.H. and S.I. Waveguide characterization and OCT imaging were performed by D.K.K. and S.T., respectively.

## Supplementary Materials

Materials and Methods

Toy Model and Numeric Simulations

Supplementary Test

Figures S1-S27

References (S1-S75)

Movies S1-S4

# Figure 1

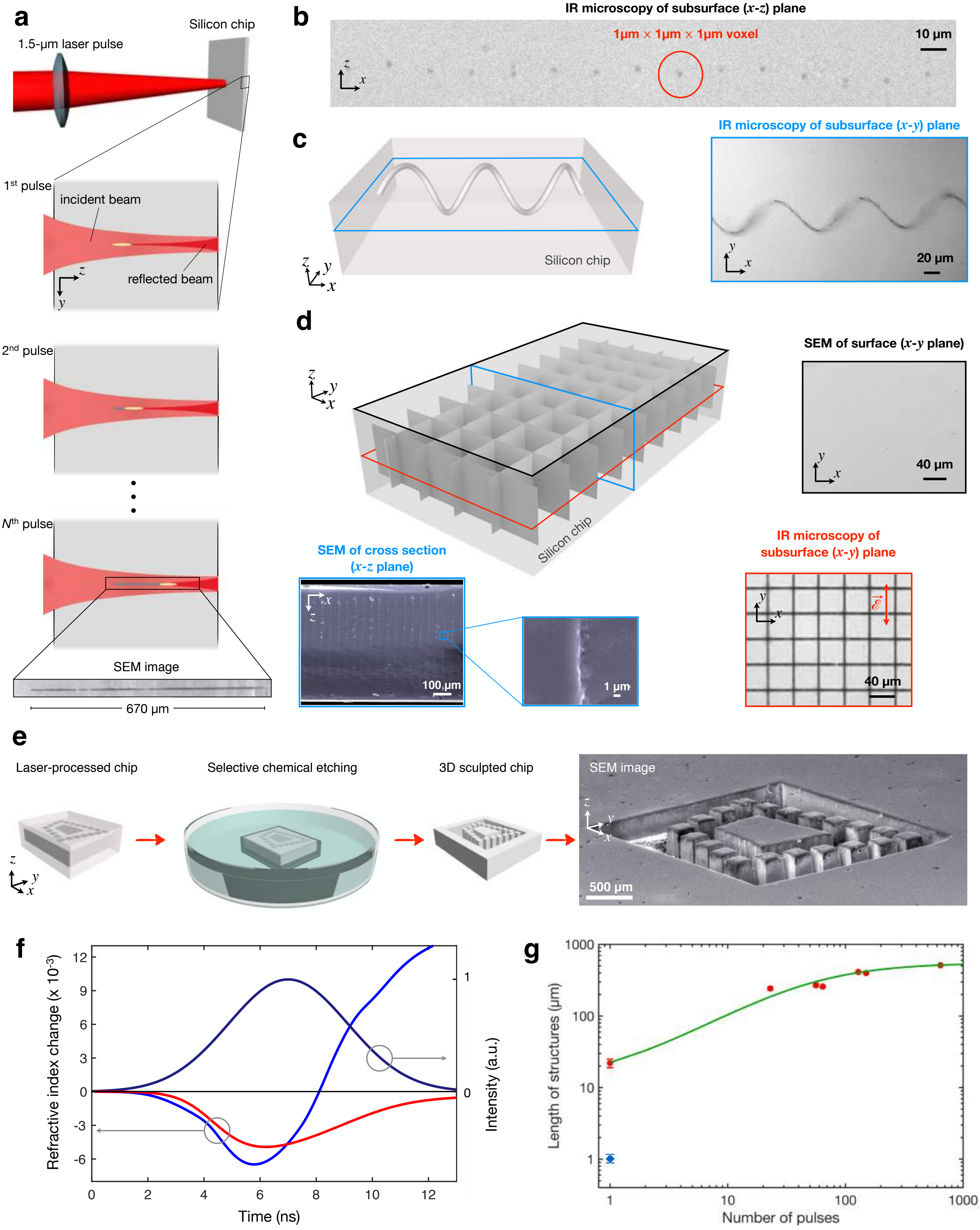

## Figure 2

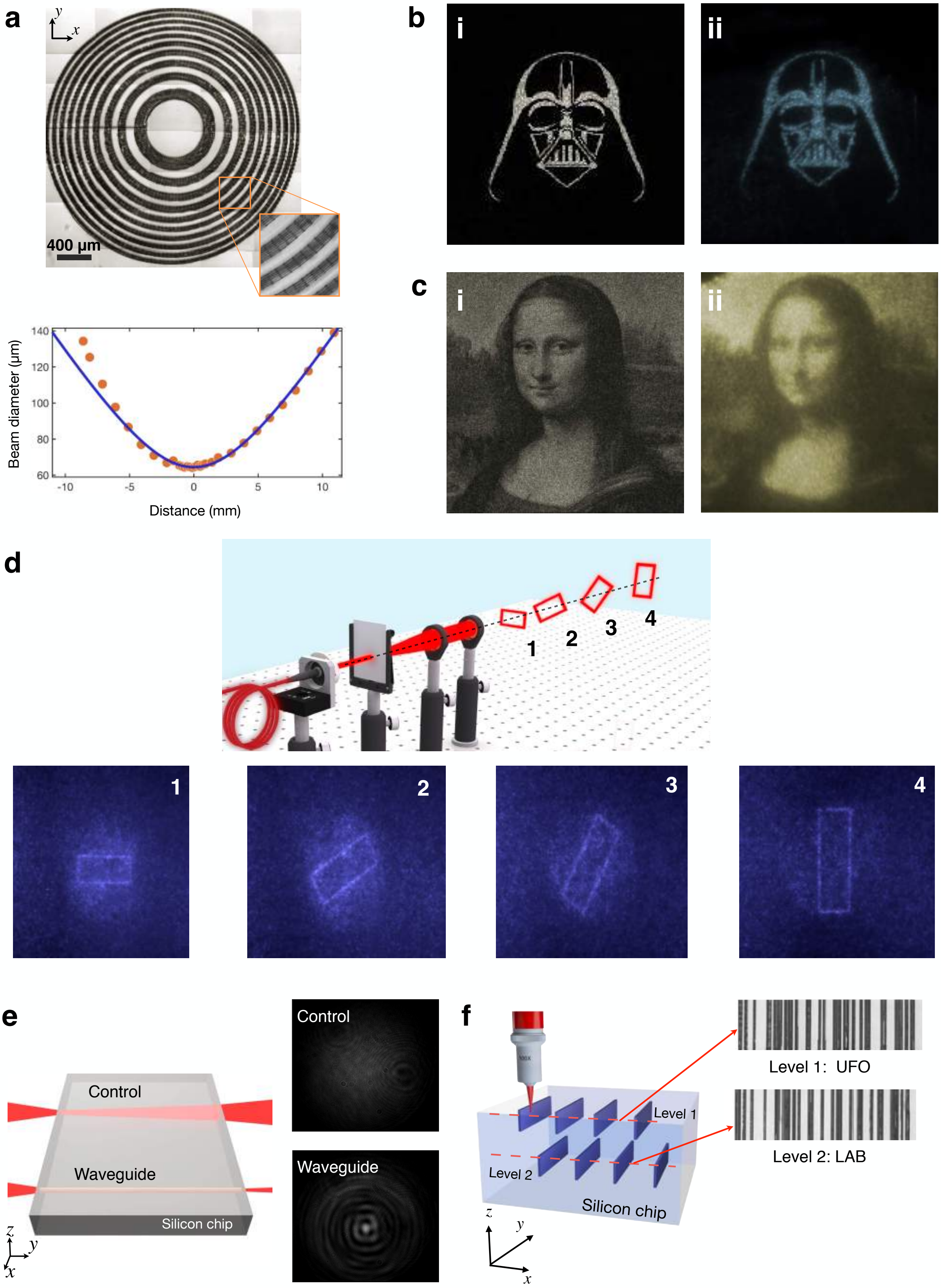

# Figure 3

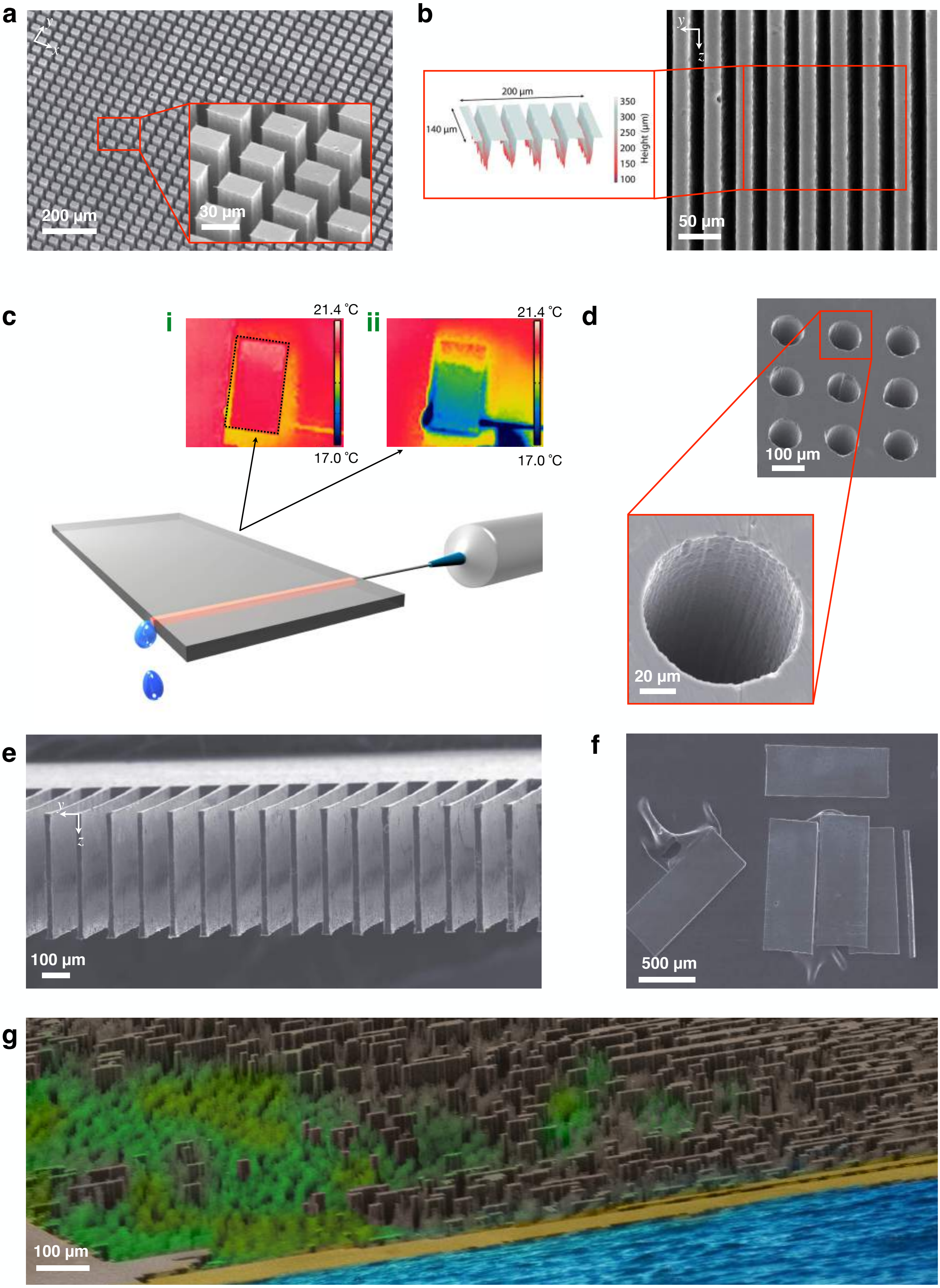